\documentstyle[aps,twocolumn]{revtex}

\def\tensor#1{{{}^{{}^{\leftrightarrow}} \!\!\!\!\!}{#1}}
\def\nbar#1{\langle #1 \rangle} \def\vnabla{\vec \nabla}
\def\rhoT{\rho^{\mbox{\tiny T}}}
\def\jT{\vec j^{\mbox{\tiny T}}}
\def\j+{\vec j^{+}}
\def\RT{R^{\mbox{\tiny T}}}
\def\sfont#1{\centerline{\sc{#1}}}

\begin{document}

\twocolumn[
\hsize\textwidth\columnwidth\hsize\csname@twocolumnfalse\endcsname
\draft

\title{Derivative relation for thermopower in the quantum Hall regime}

\author{Steven H. Simon${}^{1}$ and Nigel R. Cooper${}^2$}

\address{${}^1$Department of Physics, Massachusetts Institute of
  Technology, Cambridge, MA 02139 \\ ${}^2$Institute Laue-Langevin,
  Avenue des Martyrs, B. P. 156, 38042 Grenoble, Cedex 9, France}

\date{\today}

\maketitle

\begin{abstract}
  Recently, Tieke {\it et al} (to be published in PRL) have observed the
  relation $S_{yx} = \alpha B \, \frac{dS_{xx}}{dB}$ for the
  components of the thermopower tensor in the quantum Hall regime,
  where $\alpha$ is a constant and $B$ is the magnetic field.  Simon
  and Halperin (PRL {\bf 73}, 3278 (1994)) have suggested that an
  analogous relation observed for the resistivity tensor $R_{xx} =
  \alpha B \, \frac{d R_{xy}}{dB}$ can be explained with a model of
  classical transport in an inhomogeneous medium where the local Hall
  resistivity is a function of position and the local dissipative
  resistivity is a small constant.  In the present paper, we show that
  this new thermopower relation can be explained with a similar model.
  (This paper supersedes cond-mat/9705001 which was withdrawn)
\end{abstract}
\vspace*{10pt}

]

\setcounter{section}{1}
{\sfont{1. Introduction}}
\label{intro} \vspace{10pt}

For a wide range of conditions, high mobility quantum Hall systems
have been observed to display a derivative relation\cite{Chang}
\begin{equation}
  \label{eq:resistivitylaw}
  R_{xx} = \alpha_r B \, \frac{d R_{xy}}{dB}
\end{equation}
where $R_{xx}$ and $R_{xy}$ are the diagonal and off diagonal
components of the measured resistivity tensor $\tensor R$, $B$ is the
magnetic field, and $\alpha_r$ is a sample dependent (and weakly
temperature dependent) constant.  In Ref.~\onlinecite{Simon} an
explanation for this relation was proposed based on a classical
analysis of transport properties of a system with a local Hall
resistivity $\rho_{xy}(\vec r)$ that is a function of position and a
local longitudinal resistivity $\rho_{xx}$ which is a small constant.
It was found that if the correlations in the disorder of
$\rho_{xy}(\vec r)$ exist on several length scales\cite{Cooper}, then
the derivative law can be reasonably explained.  (A detailed
review of Ref.~\onlinecite{Simon} will be given in section
2 below).

In a recent letter, Tieke {\it et al.}\cite{Benno} have observed an
analogous derivative relation for the thermopower given by 
\begin{equation}
  \label{eq:thermopowerlaw}
  S_{yx} = \alpha_s B \, \frac{d S_{xx}}{dB}
\end{equation}
where $S_{xx}$ and $S_{yx}$ are the diagonal and off-diagonal parts of
the thermopower tensor $\tensor S$ and $\alpha_s$ is a constant found
to be approximately equal to $\alpha_r$.  (The thermopower is defined
via ${\vec{E}} = {\tensor{S}} \vec \nabla T$ under conditions where no
current is allowed to flow into or out of the sample with $\vec E$ the
electric field and $T$ the temperature).  In Ref.~\onlinecite{Benno},
it was conjectured that similar physics may be at work in thermopower
as for resistivity.  The purpose of this paper is to demonstrate that
the derivative relation for thermopower (Eq.~\ref{eq:thermopowerlaw})
can be derived in a similar manner to the derivative relation for
resistivity (Eq.~\ref{eq:resistivitylaw}).

\vspace*{10pt}
\setcounter{section}{2}
{\sfont{2. Review of Resistivity Problem}}
\vspace{10pt}
\label{sec:review}

We begin by reviewing the derivation of the derivative relation for
resistivity (Eq.~\ref{eq:resistivitylaw}) that was proposed in
Ref.~\onlinecite{Simon}.  In that work it is assumed that there
is a local resistivity tensor $\rho(\vec r)$ whose off diagonal
component $\rho_{xy}(\vec r) = -\rho_{yx}(\vec r)$ is some arbitrary
function $f$ of the local filling fraction $\nu(\vec r)$
\begin{equation}
  \label{eq:fdef}
  \rho_{xy}(\vec r) = f(\nu(\vec r))
\end{equation}
with
\begin{equation}
  \nu(\vec r) = n(\vec r) \phi_0/B
\end{equation}
where $n(\vec r)$ is the local density and $\phi_0 = h c/e$ is the
flux quantum.  The density is assumed to have some average value
$\nbar n$ and some root mean square fluctuation $\delta n$, such that
the filling fraction also has some average value $\nbar \nu$, and some
root means square fluctuation $\delta \nu$ given by
\begin{equation}
  \nbar \nu = \nbar n \phi_0/B ~~~~   ; ~~~~~ \delta \nu = \delta n \, \phi_0
  /B .
\end{equation}
(Everywhere in this paper, $\nbar {}$ is a spatial average, and
$\delta$ is a root mean square fluctuation around this average).  We
will assume that the local fluctuations in density are smooth and are
much smaller than the average density.  Thus $\rho_{xy}$ also has an
average value $\nbar {\rho_{xy}}$ and a fluctuation $\delta \rho_{xy}
\ll \nbar{\rho_{xy}}$
given by
\begin{equation}
  \nbar {\rho_{xy}} = f(\nbar \nu) ~~~~ ; ~~~~ \delta \rho_{xy} =
  \delta\nu \,|f'(\nbar \nu)| .
\end{equation}

To complete the model, we must also include a mechanism for
dissipation.  We will consider a model discussed in
Ref.~\onlinecite{Simon} which assumes that the local dissipative
resistivity $\rho_{xx} = \rho_{yy}$ is a small constant.  The major
result (Eq.~\ref{eq:resistivitylaw}) turns out to be relatively
independent of the precise model of dissipation so long as the local
dissipation is very small.  For this resistive model, one must assume
that $\rho_{xx} \ll \delta \rho_{xy}$.

In order to solve the transport problem we must satisfy current
conservation, Maxwell's equation, and Ohm's law:
\begin{eqnarray}
   \vnabla \cdot \vec j &=& 0 \label{eq:conserve} \\
   \vnabla \times \vec E &=& 0  \label{eq:Maxwell} \\
   \vec E &=& \tensor \rho \, \vec j . \label{eq:Ohm}
\end{eqnarray}
These must be supplemented with the boundary condition that a fixed
total current runs through the system, or equivalently, that the
spatial average of the current $\langle \vec j \rangle$ has a
specified value.  Substituting Eq.~\ref{eq:Ohm} into
Eq.~\ref{eq:Maxwell} and using Eq.~\ref{eq:conserve}, we obtain the
fundamental equation\cite{Simon}
\begin{equation}
  \label{eq:maineq}
  \vnabla \rho_{xy} \cdot \vec j - \rho_{xx} (\vnabla \times \vec j) = 0.
\end{equation}
This equation determines the current paths through the system and
hence the resistivity of the sample.  There are two important things
to note about this equation.  To begin with, the solution to this
equation is clearly independent of the average value of the Hall
resistivity $\nbar {\rho_{xy}}$ and can therefore only depend on its
fluctuations. Since the current profile determines the
dissipation, on dimensional grounds, in the limit of small
$\rho_{xx}$, the macroscopic dissipative resistivity $R_{xx}$ must
scale as
\begin{equation}
\label{eq:scale}
R_{xx} = C_r ( \delta {\rho}_{xy} )^{1 - \omega} \rho_{xx}^\omega
\end{equation}
with $C_r$ a dimensionless constant.  We will find below that the
exponent $\omega$ depends on the details of the disorder in the
sample, but is typically small, and can often be quite close to zero.
We will show below that a sufficiently small value of $\omega$ will
allow us to derive the derivative relation for resistivity
(Eq.~\ref{eq:resistivitylaw}).  The important physical result in
Eq.~\ref{eq:scale} is that the macroscopically measured dissipative
resistivity can depend very strongly on the fluctuations in local
$\rho_{xy}$ and can be relatively independent of the microscopic
dissipative resistivity $\rho_{xx}$.

The second thing to note about Eq.~\ref{eq:maineq} is that for
$\rho_{xx} = 0$, the current paths must flow perpendicular to the
gradient of $\rho_{xy}$, or along an equi-$\rho_{xy}$ contour.  A
nonzero $\rho_{xx}$ in Eq.~\ref{eq:maineq} can be viewed as a
diffusion constant for the current distribution\cite{Isichenko}, and
for sufficiently small $\rho_{xx}$ the current cannot diffuse very far
away from these contours.  Thus, in the limit of small $\rho_{xx}$, in
order for current to flow over distances large compared to the
correlation length of the inhomogeneities (which is assumed to be
small compared to the sample size), it must follow contours of
$\rho_{xy}$ that percolate across a macroscopic portion of the system.
We know from percolation theory\cite{Simon,Isichenko} that such a
percolating contour will be extremely convoluted.  Thus, for small
$\rho_{xx}$, the current path is anomalously long so the macroscopic
resistivity is anomalously large.

As we increase $\rho_{xx}$ two things happen.  On the one hand, the
dissipation per unit length increases, but on the other hand the
current can diffuse somewhat from the equi-$\rho_{xy}$ contours
cutting off corners of the long tortuous path, decreasing the length
of the current path, and thus acting to decrease the dissipation.
These competing effects keep the macroscopic dissipative resistivity
$R_{xx}$ relatively independent of the microscopic dissipative
resistivity $\rho_{xx}$, thus keeping the exponent $\omega$ small.
For Gaussian correlated disorder on a single length
scale\cite{Simon,Isichenko,Shklovskii} it is found that $\omega =
3/13$.  (For a similar model with viscous dissipation\cite{Simon}, one
finds $\omega = 3/19$).

If disorder exists on several length scales, however, the exponent
$\omega$ can be much smaller\cite{Simon,Isichenko}.  To see this we
consider a system where there is Gaussian correlated disorder on two
well separated length scales $l \ll l'$ which are both much less than
the size of the system.  Using the above argument we find that the
dissipative resistivity $\rho_{xx}'$ on a scale much larger than $l$
but much less than $l'$ would be $\rho_{xx}' \sim \rho_{xx}^{3/13}
(\delta \rho_{xy})^{10/13}$.  Now using $\rho_{xx}'$ as a microscopic
resistivity and repeating the argument for the disorder on length
scale $l'$ yields $R_{xx} \sim (\rho_{xx}')^{3/13} (\delta
\rho_{xy})^{10/13} \sim \rho_{xx}^{9/169} (\delta
\rho_{xy})^{160/169}$ or an exponent of $\omega = (3/13)^2$.

Throughout this work, we will assume that disorder exists on several
length scales so that the exponent $\omega$ is very small.  (The
experimental observation of the derivative relation
Eq.~\ref{eq:resistivitylaw} for resistivity will be taken as one piece
of evidence for disorder on several length scales.  Further evidence
is given in Ref.~\onlinecite{Cooper}).

We now show that a sufficiently small exponent $\omega$ results in the
derivative relation shown in Eq.~\ref{eq:resistivitylaw}.  Considering
the case of $\omega=0$, we have
\begin{equation}
\label{eq:Rxx}
  R_{xx} = C_r \delta \rho_{xy} = C_r \delta \nu \, |f'(\nbar \nu)|
\end{equation}
Note that the macroscopic dissipative resistivity here depends
entirely on the fluctuations in the microscopic $\rho_{xy}$.  On the
other hand, the macroscopic Hall resistivity is just
\begin{equation}
  \label{eq:Rxymean}
  R_{xy} = \nbar {\rho_{xy}} = f(\nbar \nu)
\end{equation}
Differentiation of this equation with respect to magnetic field (using
$d\nu/dB = -\nu/B$) leads to Eq.~\ref{eq:resistivitylaw} with
$\alpha_r = C_r \delta n/\nbar n$.  In general, we do not know the
value of $C_r$, but assuming it to be order unity yields $\alpha_r$ on
the order of a few percent which is in agreement with experimental
observation.

If the exponent $\omega$ is only slightly different from zero, then
the resistivity law Eq.~\ref{eq:resistivitylaw} will be observed to
hold to a reasonably good approximation.  If $\omega$ were
substantially different from zero, one would have to know the precise
dependence of $\rho_{xx}$ on magnetic field to make any further
statements.

\vspace*{10pt}
\setcounter{section}{3}
{\sfont{3. Mapping Thermopower To Resistivity}
\vspace{10pt}

In the case of thermopower, we will once again look for the effects of
inhomogeneities in the local transport properties on the measured
response of the sample.  Thus, we consider a local\cite{Ruzin}
thermopower tensor $\tensor{s}(\vec r)$ such that
\begin{equation}
  \label{eq:thermoohm}
  \vec E = \tensor \rho \, \vec j + \tensor s \, \vnabla T .
\end{equation}
We will write $s_{xx}(\vec r)=s_{yy}(\vec r)$ as a function $g$
of the local filling fraction, and of the magnetic field $B$,
\begin{equation}
  \label{eq:gdef}
  s_{xx}(\vec r) = g(\nu(\vec r),B) \equiv g_B(\nu(\vec r)).
\end{equation}
Note that, unlike for resistivity, we do not in general assume that
$s_{xx}$ is a function of $\nu$ only (this will be discussed further
below).  In microscopic derivations\cite{microscopic} of the
thermopower tensor $\tensor s$, appropriate for the samples studied in
Ref.~\onlinecite{Benno}, it is found that the diagonal component
$s_{xx}$ is large compared to the off diagonal component $s_{yx}$
which is small or zero.  Thus, we assume $s_{yx}=-s_{xy}$ is a small
constant (which may be zero).  Specifically, we will assume that
$s_{yx} \ll \delta s_{xx}$.  As in the case of the resistivity
problem, the precise behavior of $s_{yx}$ will not affect the outcome
of the analysis so long as it remains very small.  (In addition, our
results do not depend on whether the thermopower is dominated by the
``phonon-drag'' or ``diffusion'' contributions\cite{gallagher}).

We would like to make the thermopower problem look more like the
resistivity problem above.  To do so, we define a fictitious current
which `flows' along the isothermal lines
\begin{equation}
\jT = \hat z \times \vnabla T
\end{equation}
with $\hat z$ the unit vector normal to the plane.  Since 
\begin{equation}
  \label{eq:Tconserve}
  \vnabla \cdot \jT = \vnabla \times \vec \nabla T = 0
\end{equation}
we have $\jT$ a conserved current analogous to the charge current
$\vec j$ in the resistivity problem.  In terms of this new current,
Eq.~\ref{eq:thermoohm} is written as
\begin{equation}
  \label{eq:BigOhm}
  \vec E = \tensor \rho \, \vec j + \tensor \rhoT \,  \jT
\end{equation}
where
\begin{equation}
  \label{eq:rhoTdef}
 \tensor \rhoT(\vec r) = \left( \begin{array}{cc} \rhoT_{xx}
 & \rhoT_{xy}(\vec r)
  \\ -\rhoT_{xy}(\vec r) & \rhoT_{xx} \end{array} \right) =
 \left( \begin{array}{cc} s_{yx} & s_{xx}(\vec r)
  \\ -s_{xx}(\vec r) & s_{yx} \end{array} \right)
\end{equation}
with $\rhoT_{xx} \ll \delta \rhoT_{xy}$.

Thus, $s_{yx}$ is mapped to a dissipative resistivity $\rhoT_{xx}$
which is assumed to be a small constant, and $s_{xx}$ is mapped to a
Hall resistivity $\rhoT_{xy}$ which is a function of the local filling
fraction.  This mapping is then quite suggestive that the thermopower
law (Eq.~\ref{eq:thermopowerlaw}) might be derived analogously to the
resistivity law (Eq.~\ref{eq:resistivitylaw}).

\vspace*{10pt} \setcounter{section}{4} {\sfont{4. Thermopower Law}
  \vspace{10pt}

  In the case of thermopower for a quantum Hall sample, it is
  essential to realize that the lattice surrounding the two
  dimensional electron gas carries heat much more readily than the
  electrons (since the number of electrons in the layer is quite
  small).  The lattice surrounding the two dimensional electron gas is
  assumed to be homogeneous so that when a thermal gradient is applied
  to the sample, $\vnabla T$ is completely uniform. (Note that this
  assumes good thermal equilibration between the lattice and the
  electrons.)  Thus we should think of $\vnabla T$ (or equivalently
  $\jT$) as being applied externally to the sample and as being a
  fixed quantity which is spatially uniform.  This is very different
  from the above electrical case where only the average value $\nbar
  {\vec j}$ is fixed and the actual current distribution is quite
  inhomogeneous.  Here, we must also demand that no net electrical
  current travels in the system ($\nbar {\vec j}=0$).  By substituting
  Eq.~\ref{eq:BigOhm} into Maxwell's equation (Eq.~\ref{eq:Maxwell})
  and using current conservation (Eqs.  \ref{eq:conserve} and
  \ref{eq:Tconserve}) we obtain the fundamental equation
\begin{equation}
  \label{eq:main2}
  \vnabla \rhoT_{xy} \cdot \jT + \vnabla \rho_{xy} \cdot \vec j -
  \rho_{xx} (\vnabla \times \vec j) = 0.
\end{equation}
similar to Eq.~\ref{eq:maineq}.  Recall here that both $
\rhoT_{xy}(\vec r) = s_{xx}(\vec r) = g_B(\nu(\vec r))$ and
$\rho_{xy}(\vec r) = f(\nu(\vec r))$ are determined by the local
filling fraction.  Thus, their gradients are proportional via (see
Eqs.  \ref{eq:fdef}, \ref{eq:gdef}, and \ref{eq:rhoTdef}) $ \vnabla
\rhoT_{xy} = \gamma \vnabla \rho_{xy} $ where
\begin{equation}
  \gamma = \frac{g'_B(\nbar \nu)}{f'(\nbar \nu)} .
\end{equation}
Similarly, we have $\delta \rhoT_{xy} = \gamma \delta \rho_{xy}$.
Note that here, as in elsewhere in this work, we have assumed that
$\delta n/\nbar n$ is small enough that we need only expand quantities
linearly around the average density.

We can now define a new current
\begin{equation}
  \j+ = \vec j + \gamma \jT
\end{equation}
in terms of which the fundamental equation \ref{eq:main2} can be
rewritten as (recalling that $\jT$ is a constant)
\begin{equation}
  \vnabla \rho_{xy} \cdot \j+ - \rho_{xx} (\vnabla \times \j+) = 0 
\end{equation}
which is precisely the same as Eq.~\ref{eq:maineq}.  This must be
supplemented by the boundary condition that $ \nbar {\j+} = \nbar
{\vec j} + \gamma \nbar {\jT} = \gamma \jT$.

We thus see that current $\j+$ travels across the system in the same
inhomogeneous percolative manner as the electrical current in the
resistivity problem where $\j+$ flows only through very narrow
channels and is zero (or very small) throughout most of the volume of
the system.  Note that here $\j+$ is made up of two pieces --- a
uniform piece $\jT$ which is nonzero, and an electrical piece $\vec j$
that is highly inhomogeneous but has a zero average and carries no net
current.  The two pieces are arranged to precisely cancel throughout
most of the system and only leave a nonzero contribution to $\j+$ in
narrow channels.

Extending the analogy with the resistivity problem, we define a local
`electric' field
\begin{equation}
  \vec E^+ = \tensor \rho \, \j+,
\end{equation}
in terms of the local electrical resistivity tensor.  One can then
calculate $\vec E^+$ precisely as described in section
2 and (so long as we assume disorder on several length
scales) we obtain a macroscopic average of $\vec E^+$ that satisfies
\begin{equation}
  \nbar {\vec E^+} = \tensor R \nbar {\, \j+} = \gamma \, \tensor R \, \jT
\end{equation}
with the components of $\tensor R$ given by Eqs.~\ref{eq:Rxx} and
\ref{eq:Rxymean}.

We now calculate the actual physical electric field, by rewriting
\ref{eq:BigOhm} as
\begin{equation}
  \label{eq:newplus}
  \vec E = \vec E^+ + \vec E^-
\end{equation}
with
\begin{eqnarray}
  \label{eq:E-}
  \vec E^- &=& \tensor \rho^- \jT \\
  \tensor \rho^- &=& \tensor \rhoT - \gamma \tensor \rho.
\end{eqnarray}
Note that $\delta {\rho}^{-}_{xy} = 0$ so $\tensor \rho^-$ is a
constant tensor and  $\rho^-_{xx} \ll \gamma \delta \rho_{xy}$.

Now since $\jT$ and $\tensor \rho^-$ are both uniform in space,
Eq.~\ref{eq:E-} yields a $\vec E^-$ which is simply a constant.  We
can then write the macroscopic average of the physical electric field
as
\begin{eqnarray}
  \nbar {\vec E} &=& \tensor{\RT} \jT  \\
  \label{eq:RT}
  \tensor \RT &=& (\gamma \tensor R + \tensor{\rho^-} ).
\end{eqnarray}

For the diagonal component of
$\tensor {\RT}$, we note that since $\rho^-_{xx} \ll \gamma \delta
\rho_{xy}$, we have $\rho_{xx}^- \ll \gamma R_{xx}$ and we can neglect
$\rho^-_{xx}$ to write
\begin{eqnarray} \nonumber
  \RT_{xx} &=& \gamma R_{xx} = C_r \gamma \delta \rho_{xy} \\ &=& C_r
  \delta \rhoT_{xy} = C_r \delta \nu \, |g_B'(\nbar \nu)| .
\end{eqnarray}
For the off diagonal component of $\tensor \RT$, on the other
hand, we have
\begin{eqnarray}
  \RT_{xy} &=& \gamma R_{xy} + \nbar{\rhoT_{xy} - \gamma \rho_{xy}}
  \nonumber \\
  &=&   \nbar {\rhoT_{xy}} = g_B(\nbar \nu).
\end{eqnarray}
This result could also have been obtained by examining
Eq.~\ref{eq:BigOhm} and noting that $\nbar {\vec j}$ is fixed to be
zero.

We now convert $\tensor \RT$ back to a thermopower $\tensor S$.  Using
the macroscopic version of Eq.~\ref{eq:rhoTdef} we obtain
\begin{eqnarray}
  \label{eq:Sxyresult}
  S_{yx} &=& \RT_{xx} = C_r \delta \rhoT_{xy} = C_r \delta s_{xx} =
  C_r \delta \nu \, |g_B'(\nbar \nu)| \\ S_{xx} &=& \RT_{xy} =
  \nbar{\rhoT_{xy}} = \nbar {s_{xx}} = g_B(\nbar \nu)
  \label{eq:Sxxresult}. 
\end{eqnarray}
which are the analogous results to Eqs.~\ref{eq:Rxx}
and~\ref{eq:Rxymean}.  Note that, just as $R_{xx}$ in the resistivity
problem is independent of the small value of the local $\rho_{xx}$ and
is determined by the spatial fluctuations in $\rho_{xy}(\vec r)$,
$S_{yx}$ is independent of the small value of the local $s_{xy}$ and
is a reflection of local fluctuations in $s_{xx}(\vec r)$.  By
differentiating Eq.~\ref{eq:Sxxresult} with respect to filling
fraction and comparing with Eq.~\ref{eq:Sxyresult} we obtain
\begin{equation}
  S_{yx} = \left. - \alpha_s \nu \frac{dS_{xx}}{d\nu} \right|_B
\end{equation}
with $\alpha_s = \alpha_r = C_r \delta n/n$.  

If $s_{xx} = g(\nu,B)$ were just a function of $\nu$ this would
complete our derivation.  However, in general, this is not the case.
Thus, we write
\begin{equation}
  \label{eq:derivs}   B \frac{dS_{xx}}{dB}  = \left.  -
\nu \frac{dS_{xx}}{d\nu} \right|_{B} + \left. B
\frac{dS_{xx}}{dB} \right|_{\nu} \,\, , 
\end{equation}
and we now must assume that $S_{xx}$ (or more specifically $g(\nu,B)$)
varies strongly with $\nu$ at fixed $B$ but only slowly with $B$ at
fixed $\nu$.  This is actually quite a reasonable expectation for any
microscopic calculation \cite{microscopic}, since $S_{xx}$ oscillates
quite strongly with $\nu$.  With this assumption, we can neglect the
second term on the right of Eq.~\ref{eq:derivs} to obtain the desired
result
\begin{equation}
  S_{yx} =  \alpha_s B \, \frac{d S_{xx}}{dB} 
\end{equation}
One should note that this derivation leads to the result $\alpha_s =
\alpha_r$ which is indeed observed experimentally\cite{Benno}.

\vspace*{10pt}
\setcounter{section}{5}
{\sfont{5. Summary}
\vspace{10pt}

In this work we use a model in which the local density determines the
local Hall resistivity $\rho_{xy}(\vec r)$ as well as the local
diagonal thermopower $s_{xx}(\vec r)$.  The local dissipative
resistivity $\rho_{xx}$ as well as the off diagonal thermopower
$s_{yx}$ are assumed to be small constants such that $\rho_{xx} \ll
\delta \rho_{xy}$ and $s_{yx} \ll \delta s_{xx}$.  We also must assume
that $s_{xx}$ is a strong function of $\nu$ at fixed $B$ but a weak
function of $B$ at fixed $\nu$.  Finally, assuming that the disorder
has long range correlations, or exists on several different length
scales such that the exponent $\omega$ is close to zero, we are able
to derive the derivative relation for thermopower
(Eq.~\ref{eq:thermopowerlaw}), in close analogy with the derivation of
the corresponding law for resistivity (Eq.~\ref{eq:resistivitylaw}).

\vspace*{10pt}

The authors thank B.  Tieke and J. C. Maan for helpful discussions.
This work was supported by NSF Grant Number DMR-9523361.

\end{document}